\documentstyle[12pt,epsfig]{article}

\begin{document}
\hoffset 1.5 true cm
\voffset -0.4 true cm
\topmargin 0.0in
\evensidemargin 0.0in
\oddsidemargin 0.0in
\textwidth 6.7in
\textheight 8.7in
\parskip 10 pt
\newcommand{\rff}[1]{(\ref{#1})}
\newcommand\be{\begin{equation}}
\newcommand\nd{\end{equation}}
\newcommand\bearray{\begin{eqnarray}}
\newcommand\ndarray{\end{eqnarray}}
\newcommand{\gtsorta}{\raisebox{-0.6ex}{$\stackrel{\textstyle >}{\sim}$}} 
\newcommand{\ltsorta}{\raisebox{-0.6ex}{$\stackrel{\textstyle <}{\sim}$}} 
\renewcommand{\thefootnote}{\fnsymbol{footnote}}

\title{On Phase Transition and Self-Organized Critical State in
  Granular Packings} 
\author{Einat Aharonov$^1$, David Sparks$^2$,\\
\small $^1$ Lamont-Doherty Earth-Observatory, Columbia University, Rt. 9W, Palisades, NY\\
\small $^2$ Dept. of Geology and Geophysics, Texas A $\&$ M University, College Station, TX}

\date{\today}  \maketitle
\begin{abstract}
  We model two-dimensional systems of granular aggregates confined
  between two planes and demonstrate that at a critical grain volume
  fraction an abrupt rigidity transition occurs.  This transition is
  observed both in static and shear tests.  The grain volume fraction
  at which the transition occurs, $\nu_c$, decreases with increasing
  friction between the grains.  Densely packed grains, with a volume
  fraction $\nu> \nu_c$, display an elastic-plastic rheology.  Dilute
  packings, with $\nu<\nu_c$, display gas-like characteristics.  It is
  shown that when volume fraction is allowed to change freely (using
  constant normal stress boundary condition), it evolves
  spontaneously to $\nu_c$ under a wide range of boundary conditions,
  exhibiting 'self-organized criticality'.
\end{abstract}

Granular media is a fundamental, yet not well understood, complex
system with wide ranging applications to technological and natural
systems.  In recent years there has been much work on granular
dynamics, with emphasis on how behavior of grain aggregates may
resemble solids, liquids or gases \cite{jaeger96}.  Already Reynolds
in 1885 \cite{reynolds85} noted that loosely packed sands deform
easily as fluids while dense packings resist shear as solids.
The two phases are traditionally treated separately by 'kinetic-gas'
approaches for loosely packed grains \cite{haff83} and elasto-plastic
(often using associated plasticity \cite{vermeer90}) theories for
dense soils.  In this Letter we numerically investigate the transition
between solid and gas behaviors in granular aggregates, and show that
the phase boundary between gas and solid is also an attracting point,
to which systems naturally evolve.
We numerically model grain aggregates using a version of the popular
'discrete element method' \cite{cundall79}
which treats grains as inelastic disks with rotational and
translational degrees of freedom.  Two grains of radius $R_i$ and
$R_j$ undergo an inelastic interaction when the distance separating
them $r_{ij}$ is less than the sum of their radii. During the
interaction the
$i$th grain experiences a contact force that has both shear and normal
components: 
\be 
F_{ij}(t)=[k_n (R_i +R_j -r_{ij}) - \gamma m ({\bf
  \dot r_{ij}} \cdot {\bf \hat n})] {\bf \hat n} + [min (k_s \Delta s,
\mu ({\bf F \cdot \hat n}))] {\bf \hat s}
\label{force} 
\nd
 where ${\bf \hat n}=({\bf r}_{ij} \cdot {\bf \hat x}, {\bf r}_{ij}
\cdot {\bf \hat y})/r_{ij}$, ${\bf \hat s}=({\bf r}_{ij} \cdot {\bf
  \hat y}, -{\bf r}_{ij} \cdot {\bf \hat x})/r_{ij}$, are the unit
vectors in the normal and tangential directions respectively.
$k_n,k_s$ are the normal and shear elastic constants, $m$ is the grain
mass, $\gamma$ is a damping coefficient ensuring inelasticity of the
interaction, $\mu$ is the surface friction coefficient, and $\Delta s$
is the shear displacement since the initial contact of two particles.
Force is integrated through time to calculate grain position and
velocity.  For collisions governed by equation \rff{force} energy loss
is governed by a normal restitution coefficient $e_n=\exp(-\gamma
t_{col}/2)$, where $t_{col}=\pi (2 k_n/m -\gamma^2/4)^{-1/2}$ is the
collision time, and by frictional work which depends on the amount of
real slip (after the elastic limit is reached) and the frictional
shear force, $\mu F_n$.  Time is measured in units of undissipated
elastic wave travel time $t_0=\sqrt{m/k_n}$, and distance in units of
average disk diameter $x_0=2 \bar R$. In simulations presented here
 $k_n=1,\gamma=1, k_s=0.5$ and $m=1$.
  
\begin{figure}
\begin{center}
\includegraphics[bb= 25 316 391 516,width=4.5in]{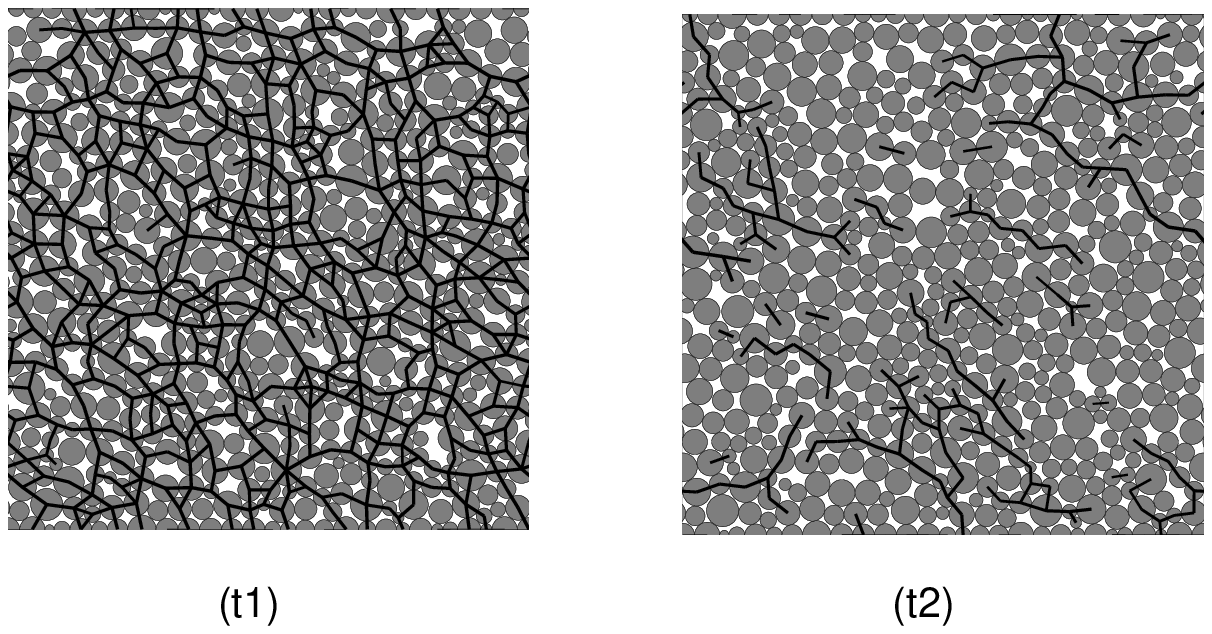}
\end{center}
\caption
{Representative instantaneous 
grain stresses and configurations during shear (here 
$v=10^{-3}$), in a CVBC, 24x24
  system, with $\mu=0.5$, at times ($t_1$) and ($t_2$).
  A line is drawn through stressed contacts. The system shown here has
  $\nu=0.80 \approx \nu_c$, and thus exhibits rigid and 'gas-like' behaviors
  intermittently during shearing: At ($t_1$), it is 'jammed' with
  system-spanning stress chain.   At time ($t_2$), the system is
  'loose', with only local stress clusters.  Systems with
  $\nu>\nu_c$ always look like the snapshot in (t1), while those with
  $\nu<\nu_c$ always look like (t2).}.
\label{chains}
\end{figure}

In this Letter we investigate three related problems. The first
problem is the general characteristics of static granular aggregates
after compaction between two parallel plates.  The second and third
problems are the behavior of the same confined aggregates during shear
using two different boundary conditions: constant distance between
shearing walls, termed constant volume boundary conditions (CVBC), and
constant normal stress applied to the wall, termed constant force
boundary conditions (CFBC).  Simulations are performed on square
systems with $n$ disks.  The top and bottom edges of the box are
composed of grains glued together to form rigid rough walls of length
$l$ (Figure \ref{chains}).
 The box is periodic in the horizontal direction.  Grain radii
are randomly drawn from a Gaussian distribution that peaks at $\bar
R$, with a standard deviation of $ 0.5 \bar R$.
Polydispersivity is introduced to discourage ordering effects.  The
system is initiated as tall loosely packed box, which is compacted
vertically by normal stresses to a height $l$.  After compaction the
horizontal walls are allowed to move freely in the horizontal
direction to relax forces. However, unrelaxed
normal forces can be maintained on the walls in this case, since we do
not allow the walls to move vertically, using CVBC.  Global
rearrangements during compaction and relaxation ensure (local) minimal
energy configuration, as would occur during natural compaction. 

\begin{figure}
\begin{center}
 \includegraphics[width=4.5in]{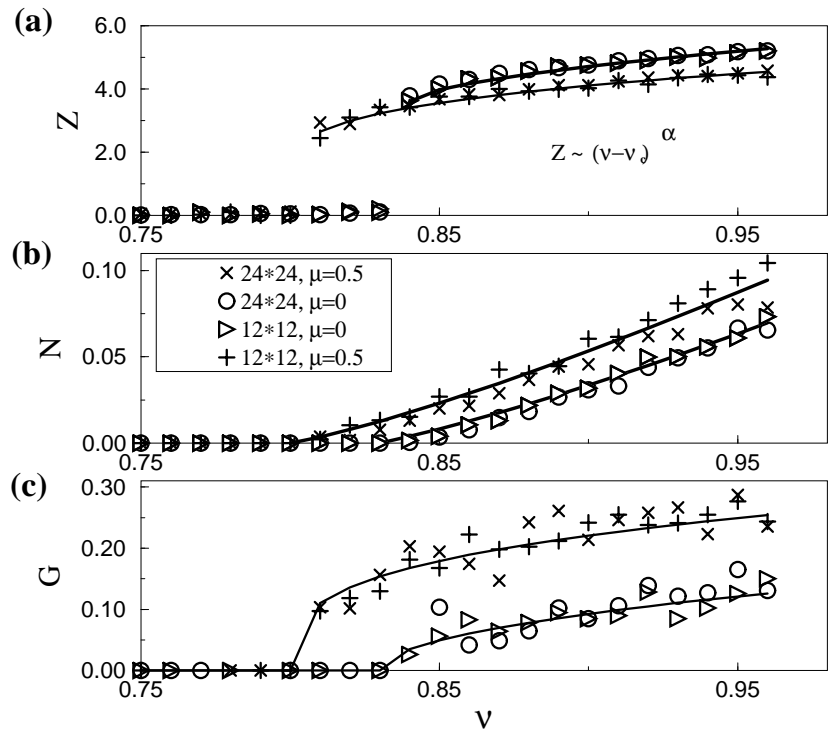}
\end{center}
\caption
{Results from CVBC simulations of static 12x12 and 24x24 grain
  packings as a function of solid fraction. (a) average
  coordination number per interior grain (b) normal stress
  exerted on horizontal walls (c) shear modulus of the
  aggregate. Solid curves are theoretical predictions}
\label{Gstep}
\end{figure}

After compaction and relaxation we measure properties of static
configurations at different solid volume fractions, $\nu=\sum_{i=1}^n
\pi R_i^2/ l^2$, ranging between 0.75 to 0.96.  Measurements of three
parameters, (a) the average number of grains touching (i.e. exerting a
force on) a grain in the interior of the box, termed the coordination
number $Z$, (b) The normal stress (normal force per unit length) $N$
operating on the upper and lower confining walls, and (c) the systems
shear modulus, G, are presented in Figure \ref{Gstep}. All measures show
an abrupt change in behavior at a critical volume fraction $\nu_c$,
which depends on the value of friction prescribed between the grains,
but  not on the system size.  The coordination number is
approximately zero for $\nu < \nu_c$. At $\nu_c$ the coordination
number abruptly jumps and then increases as an empirical  power-law, $Z \propto
(\nu-{\nu_c}_{\mu})^{\alpha_\mu}$, where the subscript $\mu$ denotes
different friction values used in simulations.  Power-law
fits (in solid curves) 
yields ${\nu_c}_{0}= 0.83 \pm 0.01,{\alpha_0}= 0.5 \pm 0.1$ and
${\nu_c}_{0.5}= 0.805 \pm 0.01,{\alpha_{0.5}} = 0.3 \pm 0.05$. 
Critical behavior with values of ${\nu_c}_0=0.82 \pm 0.02$
were obtained for hard smooth disks \cite{bideau84}, and visco-elastic 2D
bubbles \cite{durian95} for mono-dispersed and polydispersed
systems, demonstrating that $\nu_c$ is fairly independent of the disk 
size distribution, and the interaction law in absence of friction. 
The difference in $\nu_c$ and $\alpha$ 
 between frictional and smooth grains occurs 
because frictional grains tend to 'stick', and thus cannot achieve the
lower energy configuration of smooth disks.

The fact that $N$ and $Z$  (Figure \ref{Gstep}a,b) are zero for
$\nu<\nu_c$, indicates that below $\nu_c$ grains do not touch.  
At $\nu_c$ grains first touch and elastically repel each other,
exerting normal stresses on the walls.  For $\nu>\nu_c$, normal stress follows
 $N \propto Z (\nu-\nu_c)$ (shown in solid lines), a slightly modified
 form (accounting for the phase transition) of
 predictions from standard models of densely packed elastic disks
\cite{walton87}.
The observed transition is identified as a macroscopic rigidity
transition in Figure \ref{Gstep}c.  The system's elastic shear modulus
$G$ is obtained by imposing a small homogeneous shear step strain of a
magnitude $\epsilon$ and measuring the resulting shear stress on the
walls $\sigma$
($G=\sigma/\epsilon$ is independent of $\epsilon$ for $\epsilon <
10^{-4}$ of the grain radius, here we use $\epsilon=10^{-5}$).  The
procedure follows that outlined in \cite{durian95}.  Figure
\ref{Gstep}c shows $G$ increasing from zero to a finite
value, as the system passes through the phase transition at
$\nu=\nu_c$. Physically, the rigidity 
should depend on average number of springs per disk,
yet not on their compression, i.e. $G \propto Z$ (solid curves). 
The rigidity-transition thus identified is a result of global
geometrical constraints of grain packings, with the coordination number
identified as the order parameter for the transition.
 It was previously shown that $\nu=0.83 \pm 0.02$ marks
the upper limit of compacity of disordered packings of smooth hard
mono-sized and poly-sized disks \cite{bideau84}, where long-range order in
disk positions appearing for $\nu>\nu_c$.

\begin{figure}
\begin{center}
\includegraphics[ width=4.5in]{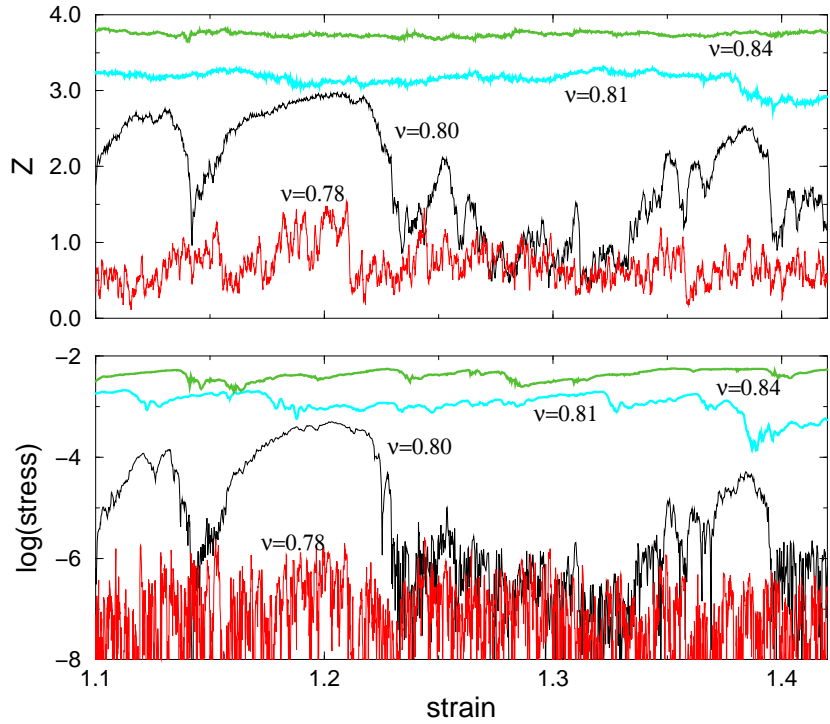}
\end{center}
\caption
{Representative grain a) coordination numbers b) shear stresses, as a
  function of strain, for a CVBC 24x24 system, with $\mu=0.5$, sheared
  at $v=10^{-3}$. Different lines represent time-series for runs having
  $\nu<\nu_c, \nu=\nu_c=0.80$, and $\nu>\nu_c$.}
\label{3phis}
\end{figure}

To investigate the transition in dynamic behavior of grains we
conducted a set of simulation shearing the static 24x24
configurations, in couette flow. Here we show results of simulations
with $\mu=0.5$. (Non-frictional grains and other values of friction
show similar behavior with the transition occuring at ${\nu_c}_{\mu}$,
and are thus not presented). The upper wall was moved at a constant
velocity ($v=10^{-3} x_0/t_0 $) while $l$ was kept constant,
maintaining CVBC.  We observe different behavior as a function of the
solid fraction: In configurations with $\nu<\nu_c$, momentum is
transferred from the wall to interior grains mostly via short-lived
collisions, observed in spiky fluctuations in stress
 measured on the wall and in coordination number, Fig. \ref{3phis}.
Stresses are transmitted only within local clusters (as in 
Figure \ref{chains}(t2)).
The power spectra of the stress fluctuations time series approaches
white noise demonstrating the uncorrelated nature of stress-transfer.
 For dense packings, with $\nu >\nu_c$, grains interact via
long-lasting contacts. Global motion is characterized by
elastic-plastic cycles: clusters of grains in contact accumulate
recoverable elastic strain, but when stresses become too  great, 
grains suddenly rearrange to relieve the
stress. Continued shearing then begins accumulation of elastic strain
on the new particle arrangement.  This behavior leads to a
stick-slip stress time series (Figure \ref{3phis}). 
 Long clusters of these grains in contact form system-spanning 'stress
chains' that transmit forces from the boundaries into the interior
(as in Figure \ref{chains}(t1)).  The stress fluctuation power spectra
for $\nu>\nu_c$ follows $f^{\eta}$, where $\eta \sim -2$, indicating
long-time correlations and in agreement with experimental results
\cite{miller96} conducted in densely packed systems. 
Most interesting is the behavior at the transition point: When
$\nu=\nu_c$ the system oscillates between a solid-like  'jammed' state
 (Figure \ref{chains}(t1)) and
 a gas-like behavior (Figure \ref{chains}(t2)).
At $\nu=\nu_c$, values of  $Z, \sigma, N$, fluctuate between
values characteristic of the 'solid-phase' and  those
characteristic of 'gas-phase' (Figure \ref{3phis}).
The transition density thus
bridges the gap between liquid and solid by coexistence of the two
phases. We note the relation to Kirkwood-Alder
transition (KAT), a disorder-order transition of repulsive 
hard-spheres studied extensively in the context of collidol
suspensions \cite{gust98}. The relation to
KAT, as well as the transition characteristics, 
suggest that the granular rigidity transition
is a first-order transition.  

The relation between stress and strain rate (to be published
elsewhere) is another property that changes across the phase boundary.
  When $\nu <\nu_c$ we measure  $(\sigma, N) \propto v^2$, as expected
from theory \cite{haff83} and experiments \cite{savage84}. 
  When $\nu> \nu_c$ measured average
stresses are nearly independent of strain rate, as expected for
elastic-plastic materials and as seen in experiments in densely packed
granular systems \cite{marone92}.  For $\nu=\nu_c$  stress-strain-rate
curves  resemble those of 
plastic materials, since high stresses in jammed states
 dominate time-averaged  behavior.

\begin{figure}
\begin{center}
  (a)\\
\includegraphics[width=3.0in]{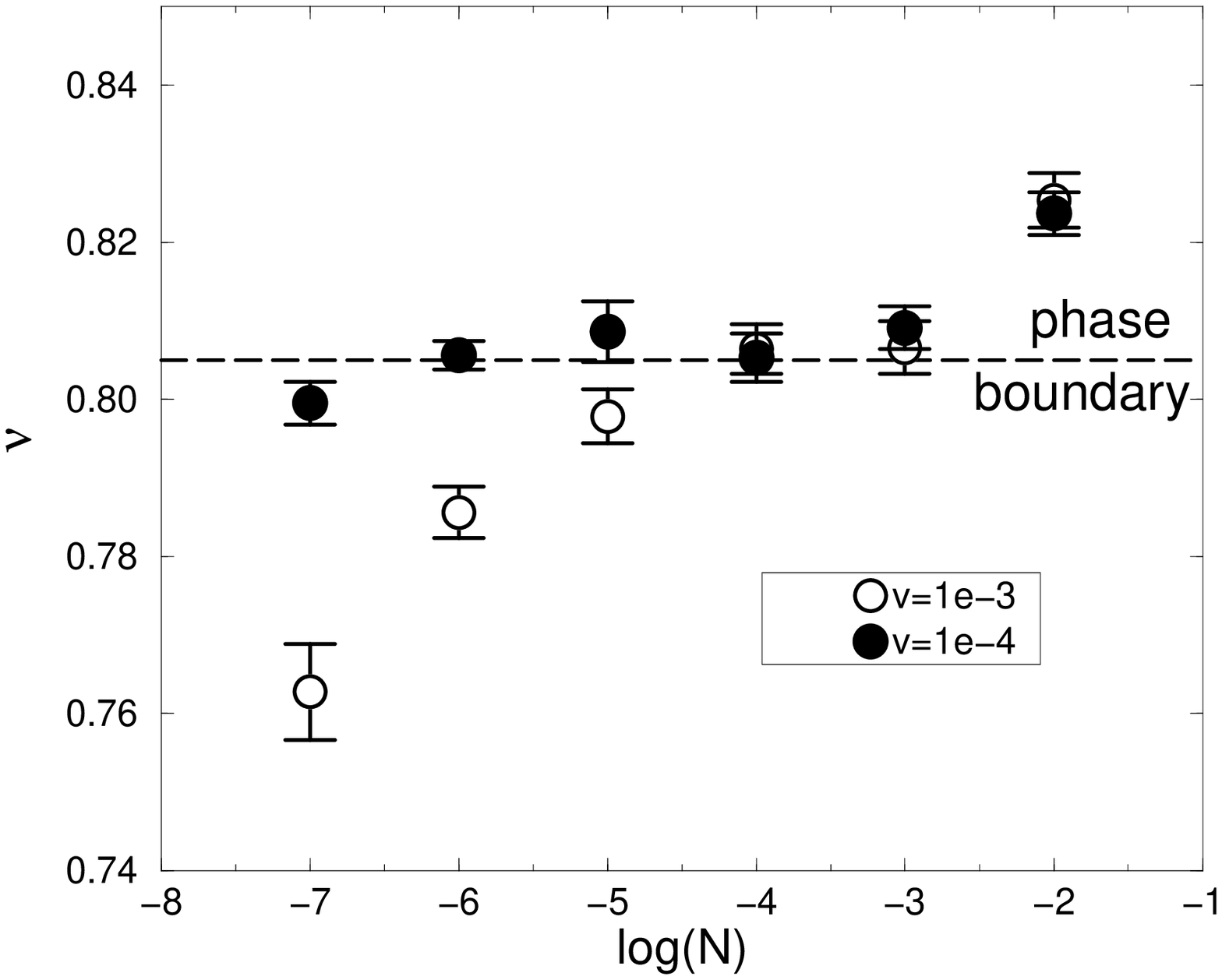}\\
 \includegraphics[bb= 90 391 445 579,width=4.5in]{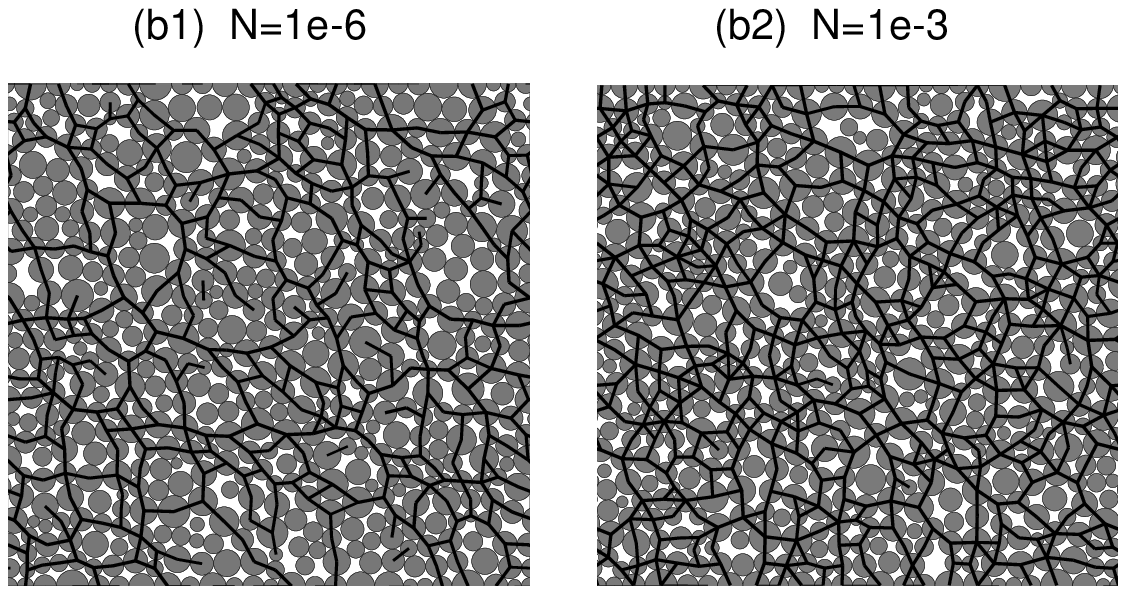}
\end{center}
\caption
{(a) Time averaged solid fraction of 24x24 systems, with $\mu=0.5$,
   sheared at two different velocities, under CFBC.  Error bars depict
  standard deviation from steady-state.
    (b) Grain configurations and stresses for two of the $v=10^{-4}$ 
runs in (a):
(b1) applied normal stress: $N=10^{-6}$,  measured time-averaged
solid fraction $<\nu>=0.806 \pm 0.002$ and 
coordination number: $<Z>=1.92 \pm 0.13$,  
(b2) $N=10^{-3}$, $<\nu>=0.809 \pm 0.003$, $<Z>=2.86 \pm 0.06$. }
\label{constN}
\end{figure}

We finally demonstrate the role of the critical solid fraction when
the system is allowed to evolve to it's own prefered solid fraction.
Soil mechanists have long known of a 'critical
density'. If a granular aggregate is over-compacted and sheared under
CFBC, it will deform while shearing and expand to this 'critical
density'. If it is initially under-consolidated it will compact while
shearing until it reaches the same 'critical density'. We perform
simulations, similar to these experimental conditions, by applying a
constant normal stresses to the rigid boundaries (CFBC), while shearing the
upper wall at velocity $v$. The systems expand or contract, depending on the
initial porosity, and finally reach a steady state where variables
($\nu, \sigma, Z$) fluctuate around a constant value. Steady
state solid fractions, $<\nu>$, are shown in Figure
\ref{constN}(a) as function of $N$ and $v$, for simulations with $\mu=0.5$. 
For a wide range of applied normal stresses
(range widens with decreasing velocity)
systems attain the critical volume fraction for
frictional grains, i.e. $<\nu> \approx {\nu_c}_{0.5}=0.805$. 
For smooth disks, simulations converge on  $<\nu> \approx
{\nu_c}_0=0.835$, in a similar manner.
Though $<\nu>$ is fairly independent of $N$
in the 'critical regime', the average coordination number $<Z>$
increases  (Figure \ref{constN}(b1)(b2)). 
In those systems having $<\nu> \approx \nu_c$
system-spanning stress-chains coexist with unstressed 'gas islands'.
With increasing $N$, the number and size of
gas islands decreases, and the connectivity of chains increases,
and thus $<Z>$ increases. Deviations from the critical state occur in
two cases: 1) At high enough normal stresses, 'solidification' occurs:
$<\nu>$ increases above $\nu_c$,
islands virtually vanish and chains become  fully connected ($N=10^{-2}$ in
Figure \ref{constN}a). 2) When the ratio of inertial to normal
forces is high inertial forces cause decompaction,
 $<\nu>$ becomes smaller than $\nu_c$
 (simulations performed at higher velocities and lower normal
stresses: Figure \ref{constN}a, $v=10^{-3}$
and $N \le 10^{-5}$), with
islands growing to divide stress chains, resulting in 'liquification'.
We also observed the  gas to solid transition in
the power spectra of stress fluctuations, where power follows
$f^{\eta}$ with $\eta $ increasing continuously from $\eta \sim 0$ for
$N=10^{-7}$ to $\eta \sim -2$ for $N \ge 10^{-2}$.  In these CFBC
simulations the system, although at the critical state, does not flip
between gas and solid phases as in CVBC, since 'jamming' episodes
may be avoided by slight dilation (producing
fluctuations around mean porosity and height). Instead of temporal
coexistence seen in CVBC the critical
state in CFBC 
is marked by spatial coexistence of two phases: gas-islands and
stress-bearing chains. 

 Why is the transition between gas and solid
also the 'critical-density' to which the system is attracted?  any
finite normal stress acting on grains would cause grain compaction and
contact, and thus loss of 'gas-like' properties, approaching $\nu_c$
from below.  Solid fractions of $\nu>\nu_c$ are characterized by
finite elastic deformation ($> 1\%$) of grains, which require
extremely high normal stresses for stiff natural granular material as
rocks.  Thus $\nu_c$ is the rigid limit, where stresses are
accommodated by efficient load bearing structures.  Experimentally, a
coexistence of liquid and solid regimes has been observed 
\cite{drake90} in shearing granular systems, and oscillations  between
'jamming' and 'flowing' states occur spontaneously in a variety
 of systems, from hoppers to natural and experimental
land-slides \cite{major97}, suggesting proximity of these systems to
the phase boundary. An attracting phase boundary 
may also explain 'fragile materials', a recent term used to describe
'jammed' states which may be unjammed by small fluctuations 
\cite{cates98}.

To summarize, we identified the rigidity transition for
granular media in $2D$ at $\nu_c= 0.80-0.84$.  Two-dimensional results can
be mapped to corresponding 3D volume fractions using a relation for
inter-particle voids, $\nu_{3D}=4 \nu_{2D}^{3/2}/3 \pi^{1/2}$
\cite{campbell85}, predicting that in 3D $\nu_{c}=0.54-0.58$ with the
lower number representing frictional grains. Experiments
\cite{hanes85} confirm that immense stiffening of rapidly shearing
grain aggregates occurs at $\nu_c \approx 0.54$.  
The critical solid-fraction is a phase-boundary
between gas and solid regimes of behavior, where the two phases were
observed to coexist in space and/or time. The critical state is
self-organizing, and corresponds to the 'critical density' known in
soil mechanics: Sheared aggregates tend to this critical volume
fraction under a wide range of normal loads.   
 We used normal stresses
ranging between $10^{-2}-10^{-7}$, which using characteristic young
modulus for Earth materials (e.g. rock), 
translates to $N=1- 10^{5}$ KPa, (corresponding for soils to burial
depth of $1-10^{5} m$). Based on this we
suggest that many natural granular deformation processes will occur
at a solid fraction which constitutes a phase boundary,
 and neither gas nor elasto-plastic
descriptions will fully capture these systems behavior.
However, such criticality in natural systems 
is easily missed, since spatial and temporal averages of stresses
 tend to be dominated by stress chains, and thus 
 have the mark of solid deformation.
It is clear that much more theoretical and experimental work needs to be
done, including investigating 
much larger systems, and examining the critical state more closely. 

\bibliographystyle{plain}
\bibliography{bibxxx}

\newpage
\newpage

\newpage

\newpage

\newpage

\end{document}